\documentclass[a4paper,11pt]{article}
\usepackage{pos}
\newcommand{\avrg}[1]{\left<  #1 \right>}

\newcommand{\ua}{\uparrow}
\newcommand{\da}{\downarrow}
\usepackage[makeroom]{cancel}

\title{Event Weighting vs. Event Counting}

\author*[a,b]{J\"org Pretz}

\affiliation[a]{Forschungszentrum J\"ulich,
  J\"ulich, Germany}

\affiliation[b]{RWTH Aachen University,
	 Aachen, Germany}

\emailAdd{pretz@physik.rwth-aachen.de}

\abstract{
	Goal of these proceedings is to introduce a method based on event weighting in particle physics experiments.
	Weighting means that events are not just counted as integer numbers
	but are assigned a weight factor according to their importance 
	in the analysis. This method has a close connection to the  maximum likelihood method known to reach the smallest statistical error.
 The purpose of this document is to give a more educational  overview on the subject.
 As an example the extraction of a beam polarization from  
 scattered particles is discussed. 
}

\FullConference{%
  RDP online PhD school and workshop "Aspects of Symmetry"(Regio2021),\\
  8-12 November 2021\\
  Online}

\begin{document}
\maketitle

\section{Introduction}
In the analysis of particle physics experiments
one often applies cuts on kinematical variables like
angles and momentum. Based on the number of event
fulfilling the cut criteria, physical 
quantities like cross section, analyzing powers or polarizations are deduced.
Aim of this work is to show that this
black and white picture, i.e.
counting an event inside the cuts and rejecting it
otherwise does in general not lead
to optimal results in terms of statistical accuracy.
It is shown that one can do better
by assigning a weight factor to every event.
A weight factor 0 corresponds to ignoring this event,
a weight factor of 1 corresponds to counting the event.
But an event could have a arbitrary weight factor between 
0 and 1. This makes better use of information an event carries.
How this is done quantitatively is shown in an educational
way for the determination of particle beam polarization.
This work is largely based on reference~\cite{Pretz:2018bze}.

Many more examples using event weighting can be found in the literature,
like the determination of the anomalous magnetic moment of the muon
\cite{Muong-2:2021ojo,Muong-2:2007tpa}, the gluon polarisation in the nucleon~\cite{Alekseev:2008cz,Pretz:2008mi}, the polarized quark distribution in the nucleon~\cite{Pretz:2016mjk}, analyzing power in deuteron carbon scattering~\cite{Muller:2020anv}, three gauge boson coupling~\cite{Diehl:1993br}, electric or magnetic dipole moment
of the $\tau$-lepton~\cite{Atwood:1991ka,Bernreuther:2021elu} and
the measurement of the forward-backward asymmetry of Drell-Yan dilepton pairs~\cite{Bodek:2010qg}.
In this context the term optimal observable is widely used in the literature.

This document is organized as follows. 
First a short introduction is given how event rates relate
to a beam polarization. Then various methods to extract the beam polarization are discussed and compared.

\section{Formulation of the problem}
Starting point of experiments in particle physics is often
the relation between the expected number of observed events $N$ and the cross section $\sigma$ containing the physics of the scattering process:
\begin{equation}\label{eq:Nunpol}
 \avrg{\frac{\mathrm{d} N(\vartheta,\varphi)}{\mathrm{d}\Omega} }  =  {\cal L} a(\vartheta,\varphi) \frac{\mathrm{d} \sigma(\vartheta)}{\mathrm{d}\Omega} \, .
\end{equation}

\begin{table}[hpb]
	\caption{Definitions of variables used in equations~\ref{eq:Nunpol} and \ref{eq:Npol}.\label{tab:var}}
	\begin{center}
		\begin{tabular}{|l|l|}
			\hline
			variable & meaning \\
			\hline
			$N(\vartheta,\varphi)$ & number of scattered particles observed \\
			$\avrg{N(\vartheta,\varphi)}$ & expectation value of number of events \\
			$\sigma(\vartheta)$ & unpolarized cross section \\
			$\mathcal{L}$ & luminosity \\
			$a(\vartheta,\varphi)$ & acceptance/efficiency \\
			$\mathrm{d} \Omega$  & solid angle \\
			$\vartheta$ & polar angle \\
			$\varphi$ & azimuthal angle, $\varphi=0$ corresponds to positive $x$-direction \\    
			$P$  & beam polarization \\
			$A(\vartheta)$  & analyzing power of scattering process \\
			\hline
		\end{tabular}
	\end{center}
\end{table}

\begin{figure}[h]
	\begin{center}
\includegraphics[width=0.6\textwidth]{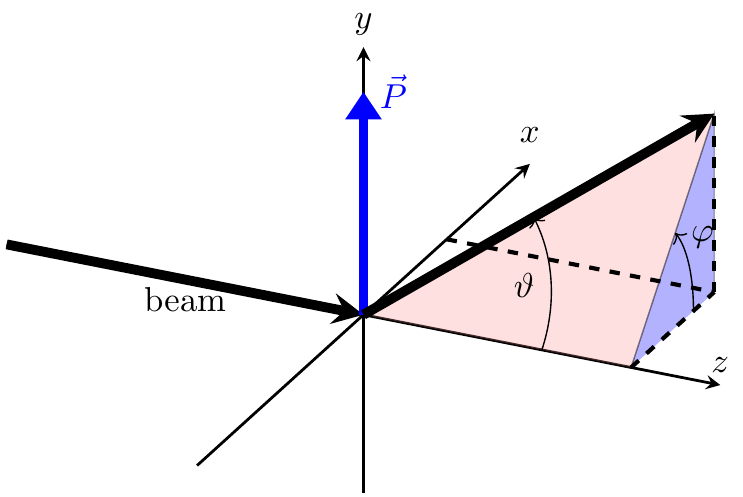}
\end{center}
\caption{Definition of polar and azimuthal angles $\vartheta$ and $\varphi$ and direction of the beam polarization vector. The target is located at the origin of the coodinate system.\label{fig:coord}}
\end{figure}

Table~\ref{tab:var} and figure~\ref{fig:coord} 
explain the variables and the coordinate system. Equation~\ref{eq:Nunpol} holds for an
unpolarized beam and target. If the beam is polarized additional contributions appear.
Here we assume a spin polarized particle beam perpendicular to the momentum vector in
$y$-direction.
In this case the cross section receives additional contributions~\cite{Ohlsen:1973wf}:
\begin{equation}\label{eq:Npol}
 \avrg{\frac{\mathrm{d} N(\vartheta,\varphi)}{\mathrm{d}\Omega}} =  {\cal L} a(\vartheta,\varphi) \frac{\mathrm{d}\sigma(\vartheta)}{\mathrm{d}\Omega} \left(1 + A(\vartheta) P \cos(\varphi)\right) \, .
\end{equation}
A detailed derivation is given in appendix~\ref{app:N2crosssection}.

For this paper the most interesting variables in equation~\ref{eq:Npol}
are the beam polarisation $P$, describing the degree of polarization of the beam and the analyzing power $A$ describing how much the scattering process distinguishes between particles scattered to the left and right. 
As one can see from equation~\ref{eq:Npol}, one is only sensitive
to the product of the analyzing power $A$ and the polarization $P$.
Here we assume that the analyzing power is known and the goal
is to extract from the data, i.e. the measured event kinematics, the polarization $P$, with the smallest possible statistical error.

\begin{figure}
	\begin{center}
		\includegraphics[width=0.6\textwidth,page=1]{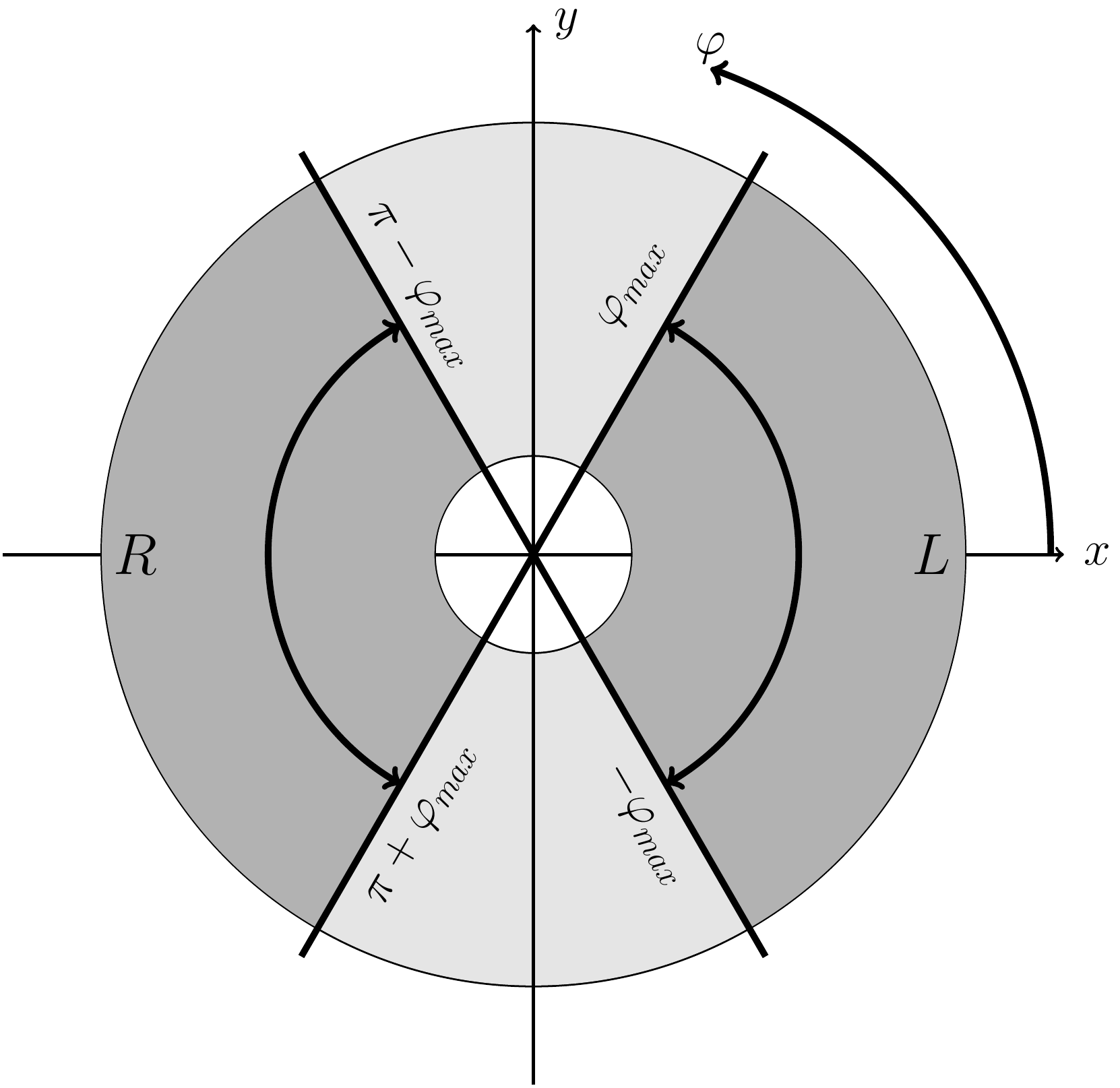}
	\end{center}
	\caption{Events are accepted in the range 
		$-\varphi_{max} < \varphi < \varphi_{max}$ (events scattered to the left) and $\pi+\varphi_{max} < \varphi < \pi-\varphi_{max}$ (events scattered to the right).
		The particle beam comes out of the plane.
		\label{fig:phi_acc}}
\end{figure}

This seems to be an easy task if all other quantities in equation~\ref{eq:Npol} are known. This is often not the case.
The cross section, the luminosities and the acceptance are in general not known or at least not to the required precision.
Several methods to extract $P$ are discussed in the following.
These methods make use of ratios where these unknown factors drop out.
To simplify the discussion we consider a fixed polar angle $\vartheta$
and an acceptance constant in $\varphi$\footnote{
In reality this is not the case. How to extract $P$ in this case 
is discussed in~\cite{Pretz:2018bze}.}.
In this case the events are distributed according to
\begin{equation}\label{eq:Nphi}
\avrg{\frac{\mathrm{d} N(\varphi)}{\mathrm{d}\varphi}} = \frac{N_0}{2 \pi} (1+A P \cos(\varphi))
\end{equation}
with $N_0 =  {\cal L}  a \int \frac{\mathrm{d}\sigma(\vartheta)}{\mathrm{d}\Omega} \mathrm{d}\Omega = {\cal L}  a \sigma$.
If we restrict the acceptance to the dark grey area in figure~\ref{fig:phi_acc} the probability density function (pdf) is given by: 
\begin{equation}
p(\varphi) = \frac{1}{4\varphi_{max}} \left(1 + AP\cos(\varphi)  \right) \, 
\end{equation}

\section{Different ways to extract $P$}
\subsection{Determing $P$ using counting rates}

Events are counted in the region between
$-\varphi_{max} < \varphi < \varphi_{max}$ 
and $\pi-\varphi_{max} < \varphi < \pi+\varphi_{max}$ 
as indicated in figure~\ref{fig:phi_acc}.
The expectation values for the number of events in the left
part of the detector is given by
\begin{eqnarray}
  \avrg{N_L} &=& \frac{N_0}{2\pi} \, \int_{-\varphi_{max}}^{\varphi_{max}}
  \left( 1+ AP \cos(\varphi)\right) \mathrm{d} \varphi \nonumber \\
  &=& \frac{N_0}{2\pi} \, 
  \left( 2\varphi_{max} + AP  \int_{-\varphi_{max}}^{\varphi_{max}}\cos(\varphi)\right) \mathrm{d} \varphi   \nonumber\\
   &=& \frac{N_0}{\pi} \varphi_{max} \,
 \left( 1 + AP  \frac{\int_{-\varphi_{max}}^{\varphi_{max}} \cos(\varphi) \mathrm{d} \varphi }{2\varphi_{max}} \right)  \nonumber\\
 &=& \frac{N_0}{\pi} \varphi_{max} \, \left( 1 + AP \,  \overline{\cos(\varphi)} \right) \, .
\end{eqnarray}
In a similar way one finds
\begin{equation}
\avrg{N_R} = \frac{N_0}{\pi} \varphi_{max} \, \left( 1 - AP \, \overline{\cos(\varphi)} \right) \, .
\end{equation}
The following estimator can be used to determine $P$:
\begin{equation}\label{eq:Pcnt}
\hat P = \frac{1}{A\, \overline{\cos(\varphi)}} \, \frac{N_L - N_R}{N_L + N_R} \, .
\end{equation}
$N_L$ and $N_R$ are the actual measured number of events.
In this ratio $a$,$\mathcal{L}$ and $\sigma$ hidden in $N_0$ drop out.
The average $\overline{\cos(\varphi)}$ can be evaluated directly from the data, i.e. 
\begin{eqnarray}
\overline{\cos(\varphi)} &=&  \frac{\int_{-\varphi_{max}}^{\varphi_{max}} \cos(\varphi) \mathrm{d} \varphi }{2\varphi_{max}} =  -\frac{\int_{\pi-\varphi_{max}}^{\pi+\varphi_{max}} \cos(\varphi) \mathrm{d} \varphi }{2\varphi_{max}} \nonumber \\ 
&=&   \frac{\sum_L \cos(\varphi_i) -\sum_R \cos(\varphi_i)}{N_L + N_R}   \, .\label{eq:avgcos}
\end{eqnarray}
Throughout the paper we assume that the number
of events is large enough such that expectation values 
or averages like in equation~\ref{eq:avgcos}
can be replaced to the corresponding sums over the event sample.

Simple Gaussian error propagation for equation~\ref{eq:Pcnt} leads to a statistical error for $P$, assuming $PA \ll 1$,
\begin{equation}
\sigma^2_{\hat P} = \frac{1}{\left(A \overline{\cos(\varphi)}\right)^2 N} \, ,
\end{equation}
where $N=N_L + N_R$ is the total number of events.

\begin{figure}
	\begin{center}
	\includegraphics[width=0.8\textwidth]{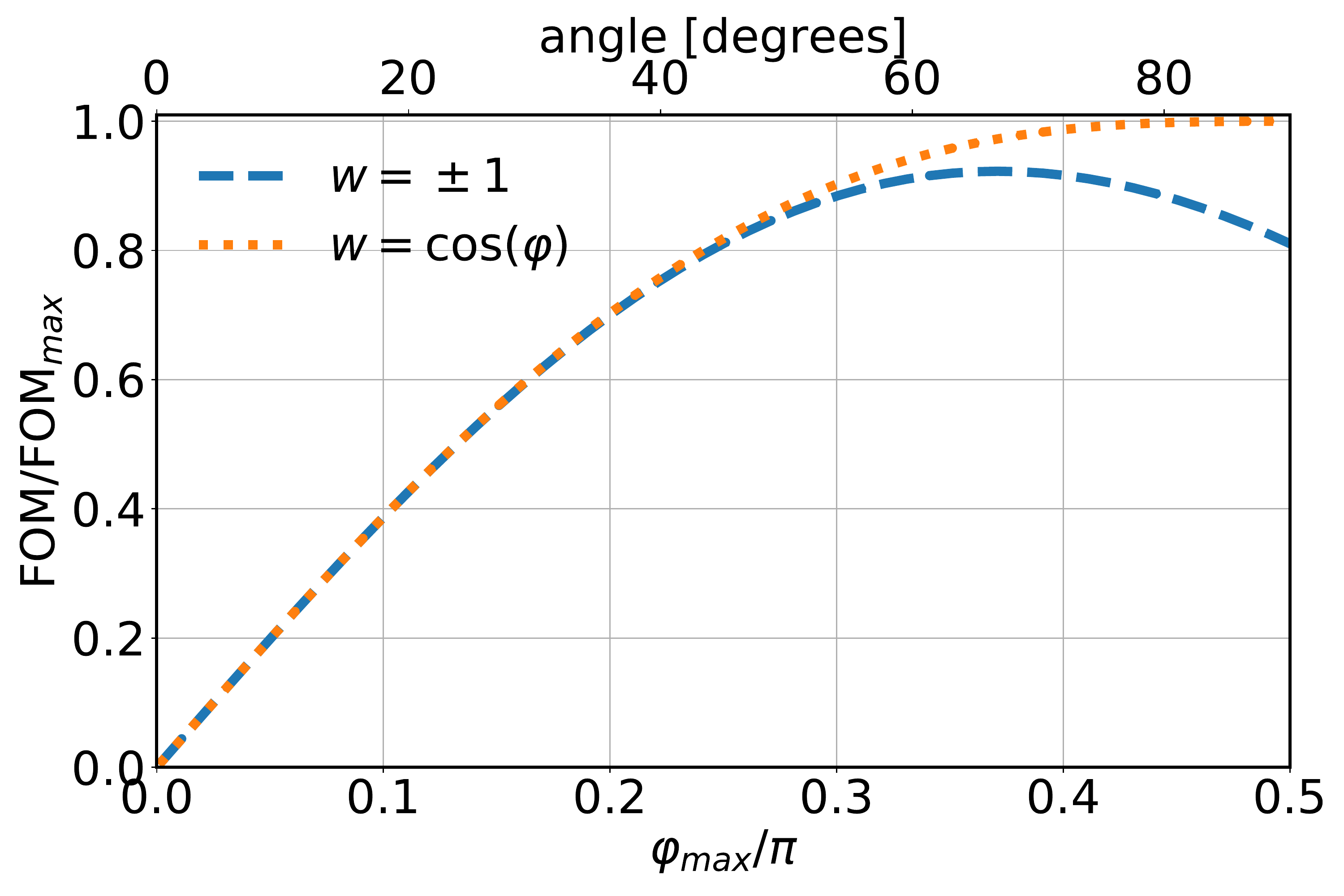}
	\end{center}
	\caption{Figure of merit (FOM) as a function of $\varphi_{max}$ for the two weighting factors. \label{fig:fom_1}}
\end{figure}

In this context it is convenient to define the figure of merit (FOM)
as
\begin{equation}\label{eq:fom_cnt}
\mbox{FOM} = \sigma^{-2}_{\hat P} =\left( A \overline{\cos(\varphi)}\right)^2 N
\end{equation} 
The dashed curve in figure~\ref{fig:fom_1} shows the FOM as a function of $\varphi_{max}$.
The curve shows an interesting behavior. Increasing
$\varphi_{max}$ the FOM increases as one would expect by adding more and more events. But from 67 degrees on the FOM decreases. Thus by adding events
the statistical error decreases, indicating that this is not the optimal way to analyze the data. The reason is the following. Increasing $\varphi_{max}$ one adds more data where $|\cos(\vartheta)|$   is small. Looking at equation~\ref{eq:Npol}
one realizes that the smaller $|\cos(\vartheta)|$ the less sensitive one is to $P$.
Adding these events dilutes the sample and leads to an increase of the statistical error on $\hat P$. 
The following section discusses a more efficient way to analyze the data.

\subsection{Determining $P$ using event weighting}\label{sec:weight}
Now we consider the following estimator
\begin{equation}\label{eq:Pw}
 \hat P = \frac{1}{A} \, \frac{\sum_{L,R} w(\varphi_i)}{\sum_{L,R} w(\varphi_i) \cos(\varphi_i)} \, .
\end{equation}
Here every event is assigned a, for the moment almost arbitrary, weight factor $w(\varphi)$.
The only constraint is that $w(\varphi) = -w(\varphi+\pi)$. In this case equation~\ref{eq:Pw} provides
an unbiased estimator for $P$.
This can be verified by looking at the expectation values
of the sums in  equation~\ref{eq:Pw}:
\begin{eqnarray}
AP \avrg{\sum_i w(\varphi_i) \cos(\varphi_i)}& =& \avrg{\sum w(\varphi_i)} \,  \nonumber \\
AP N \int_{acc} w(\varphi_i) \cos(\varphi_i) p(\varphi) \mathrm{d} \varphi&=& N \int_{acc} w(\varphi_i) p(\varphi ) \mathrm{d} \varphi  \,  \nonumber \\
AP  \int_{acc} w(\varphi_i) \cos(\varphi_i) (1+\xcancel{PA\cos(\varphi)}) \mathrm{d} \varphi&=&  \int_{acc} w(\varphi_i) (\xcancel{1}+PA\cos(\varphi))\mathrm{d}  \varphi      \, . \label{eq:expec}
\end{eqnarray}

Since the cosine function has the same properties as $w$,  $\cos(\varphi)= - \cos(\varphi+\pi))$, the crossed out terms in equation
\ref{eq:expec} vanish, leaving on both sides $AP \int w(\varphi) \cos(\varphi) \mathrm{d}\varphi = 4 \varphi_{max} \overline{w \cos(\varphi)}$.
Note that there is a subtle difference between average denoted
as $\overline{(\dots)}$ and expectation values $\avrg{\dots}$
discussed further in appendix~\ref{app:avrg}.

\begin{figure}
	\includegraphics[width=\textwidth,page=2]{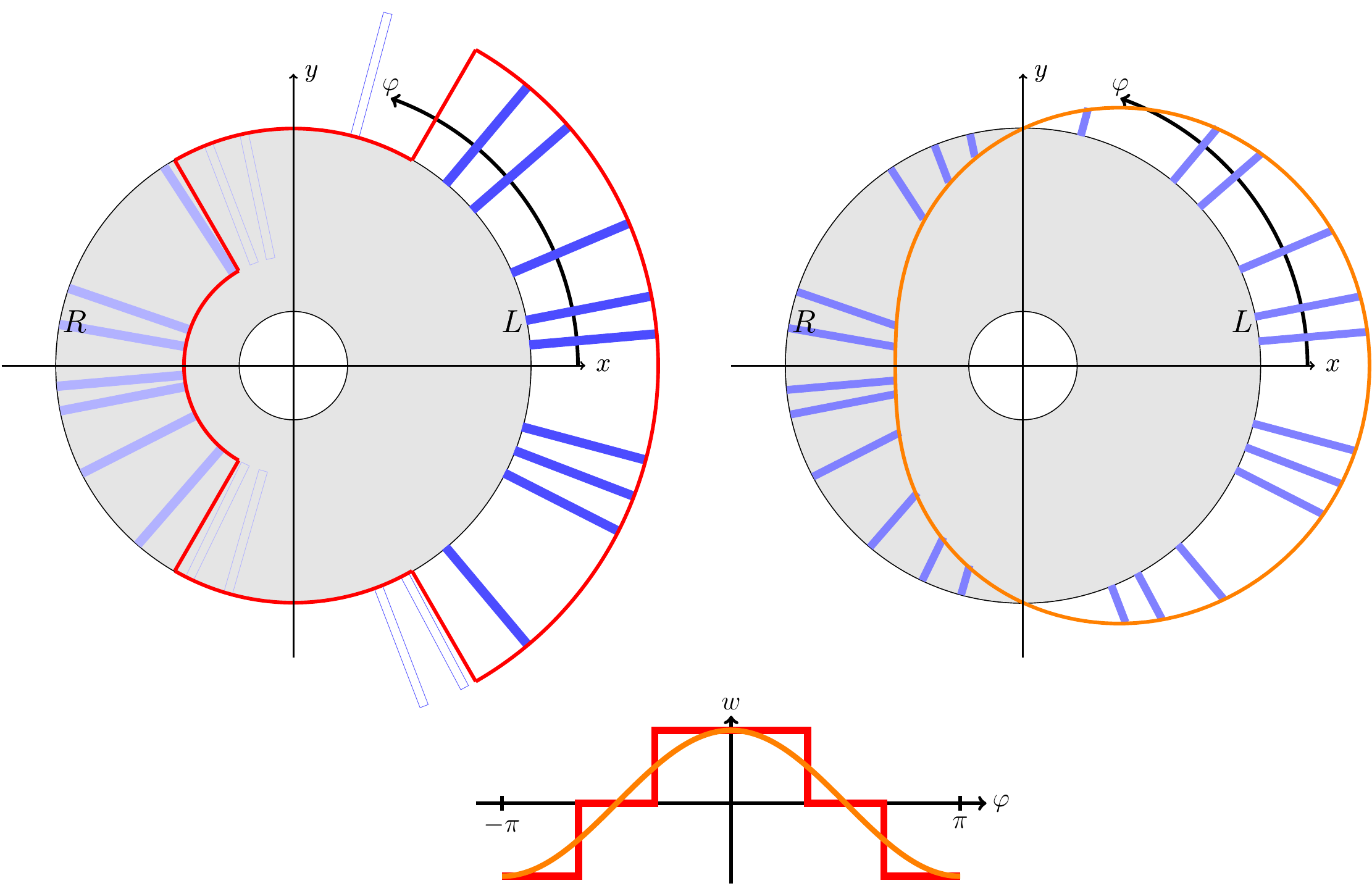}
	\caption{Left: event counting, Right: Weighting events with $\cos(\varphi)$. 
		In case of event counting events are only accepted in the indicated $\varphi$ range. For the weighting method no cut is applied. The graph at the bottom shows the two weighting factors as a function of $\varphi$.\label{fig:weight}}
\end{figure}

Two cases are of interest
\begin{itemize}
	\item simple event counting:\\
	$w = 
	\begin{cases}
	\, \, \,1, & \text{for $-\varphi_{max} < \varphi < \varphi_{max}$}\\
	-1, & \text{for $\pi-\varphi_{max}<\varphi < \pi+\varphi_{max}$}\\
	\, \,\,0, & \text{otherwise}\\
	\end{cases} $

	\item event weighting:  \\
	 $w=
	\begin{cases}
	 \cos(\varphi), &\text{for $-\varphi_{max} < \varphi <\varphi_{max}$ and $\pi-\varphi_{max}<\varphi < \pi+\varphi_{max}$} \\
	 	0, & \text{otherwise}\\
	 	\end{cases}$
\end{itemize}
The two cases are indicated in a pictorial way
in figure~\ref{fig:weight}.
Note that in the case $w=\pm1$ the estimator in equation~\ref{eq:Pw}
coincides with equation~\ref{eq:Pcnt}, if $\overline{\cos{\varphi}}$
is evaluated from data (see equation~\ref{eq:avgcos}).

Although both cases lead to an unbiased estimator for $P$,
they lead to different statistical accuracy. This will be discussed
in the next subsection.

\subsection{Comparison of the two weighting methods}
The statistical accuracy or FOM can be obtained by simple error propagation starting from equation~\ref{eq:Pw}. Details of the calculations are given in 
appendix~\ref{app:err}. Here we just quote the result
(again neglecting contributions of order $(AP)^2$ and higher) 
\begin{eqnarray}
\mbox{FOM}^{\pm 1} &=& N A^2 \, \overline{\cos(\varphi)}^2  \, ,\label{eq:fom_1}\\
\mbox{FOM}^{\cos(\varphi)} &=& N A^2 \, \overline{\cos(\varphi)^2} \label{eq:fom_c} \, .
\end{eqnarray}
The ratio of the two FOMs is
\[
\frac{\overline{\cos(\varphi)^2}}{\left(\overline{\cos(\varphi)}\right)^2} \ge 1 \, .
\]
I.e. weighting the events with $\cos(\vartheta)$ leads
in general to a higher FOM compared to the counting rate 
asymmetry ($w=\pm1$).
The comparison of the FOMs is shown in figure~\ref{fig:fom_1}
by the solid and the dashed curve.
At small $\varphi_{max}$ the difference between the two methods is
very small. From $\varphi_{max} \approx 67$~degrees on the FOM of the counting
method starts to decrease because the sample is diluted
by events at small $\cos(\varphi$). Using $w=\cos(\varphi)$
the FOM still increases.
This means that one can include
all events without decreasing the FOM. Events at $\varphi = \pm \pi/2$ receive
a weight factor 0 and are thus effectively not counted because they don't
carry any information about the polarization.
This is also reflected in the fact that the dotted curve
in figure~\ref{fig:fom_1} has slope zero at $\varphi=\pi/2$.
In the counting method one should apply a cut at $\varphi_{max }\approx 67$~degrees in order to maximize the FOM.
Comparing the maxima of both curves, the gain in FOM is about 8\%.

In the next subsection we show that the choice $w=\cos(\varphi)$ leads to the largest reachable FOM.

\subsection{Connection to Maximum Likelihood Estimator}
In this subsection it is shown, that the maximum likelihood method
leads to same estimator for $\hat P$ as the weighting method with $w=\cos(\varphi)$. Since the maximum likelihood estimator is known,
at least in the large $N$ limit, to reach the minimum variance bound,
this also proves that by weighting with $w=\cos(\varphi)$ one reaches
the smallest statistical error or the largest FOM.

The log-likelihood function for the event distribution in equation~\ref{eq:Npol} reads
\begin{equation}
  \ell = \frac{1}{4 \varphi_{max}} \, \sum_{i=1}^{N} \ln \left( 1 +  AP \cos(\varphi) \right) \, .
\end{equation}
Since the number of events $N$ is not fixed here, one should use
the extended maximum likelihood method. Since $N$ does not depend
on the parameter $P$ to be estimated, this is not important here.

The likelihood estimator is derived from
\[
 \frac{\partial \ell}{\partial P}  = \sum_i \frac{ A\cos(\varphi_i) }{1+A P\cos(\varphi_i) } \stackrel{!}{=}0 \, .
\]
Assuming $A P \ll 1$, one arrives at an analytic expression 
\[
\hat P = \frac{ \sum_{i} A \cos(\varphi)}{\sum_{i} (A\cos(\varphi_i))^2} = 
	\frac{1}{A} \, \frac{ \sum_{i} \cos(\varphi_i)}{ \sum_{i}  \cos^2(\varphi_i)}  \, ,
\]
which is identical to the expression for $w=\cos(\varphi)$
in equation~\ref{eq:Pw}.

Concerning the FOM, one finds
\begin{eqnarray}
	\mbox{FOM}^{LH} &=& -\avrg{\frac{\partial^2 \ell}{\partial P^2} }  \nonumber\\
	                &=& \avrg{\frac{A^2 \cos(\varphi)^2}{(1+AP\cos(\varphi))^2}}  \nonumber \\
                    &=& \frac{N}{4\varphi_{max}} \int \frac{A^2 \cos(\varphi)^2}{1+AP\cos(\varphi)} \mathrm{d}\varphi  \nonumber\\
                     &\approx& \frac{N}{4\varphi_{max}} \int A^2 \cos(\varphi)^2 \left( 1-AP\cos(\varphi) + (AP\cos(\varphi))^2  \right)\mathrm{d}\varphi \nonumber \\
                     &=&  NA^2 \left( 
                     \overline{\cos(\varphi)^2}  +
                     (AP)^2   \overline{\cos(\varphi)^4} \right) \nonumber   \\ 
                     &=& N A^2 \, \overline{\cos(\varphi)^2}\left(
                     1+ (AP)^2 \, \frac{\,\overline{\cos(\varphi)^4}\,}{\overline{\, \cos(\varphi)^2}\,} 
                       \right) \, ,\label{eq:fom_ml}
\end{eqnarray}
which agrees with the FOM derived for $w=\cos(\varphi)$
in appendix~\ref{app:err} using error propagation,
if one uses the approximation $(1+\epsilon) \approx (1-\epsilon)^{-1}$.






\section{Summary and Outlook}
In this note it has been shown that simply counting events is 
not always the best way to analyze data. One can minimize the statistical error on a polarization measurement by assigning
to every event an appropriate weight factor.
Concerning the statistical accuracy it was shown that a weighting
factor can be found which leads to the same estimator as 
the maximum likelihood method, known
to reach the smallest statistical error. There are cases 
where the likelihood may not be applied directly, e.g. when the
pdf is not completely known like in case of unknown
acceptance and luminosity factor. In this cases an estimator
using event weighting can often still be found~\cite{Alekseev:2008cz,Muller:2020anv}.

The method can be generalized. If the detector acceptance 
is not restricted to one value of $\vartheta$, the analyzing power $A(\vartheta)$ can
also be included in the optimal weight factor:
\[
w = A(\vartheta) \cos(\varphi) \, .
\]
In general, if events are distributed according to
\[
 p(\vec x) \propto (1 + \beta(\vec x) P) \, 
\]
where $\vec x$ denotes a set of variables, the optimal
weight factor is $w=\beta(\vec x)$ at least for $\beta P \ll 1$.
The optimal weight for the arbitrary  $\beta P$ is discussed in~\cite{Pretz:2011qb}. 

\newpage
\appendix
\section{From event rates to cross section}\label{app:N2crosssection}

This appendix shows the derivation of equation
\begin{equation}
N(\vartheta,\varphi)=  a(\vartheta,\varphi) {\cal L} \sigma(\vartheta)
\left( 1 + P A(\vartheta) \cos(\varphi)\right)
\end{equation}

The number of particles scattered to the left and right are  given by
\begin{eqnarray}
		N(\varphi=0)& = &a \rho \ell \left(n^{\ua} \sigma_{\ua,L} + n^{\da} \sigma_{\da,L} \right) =  a \rho \ell \left(n^{\ua} \sigma_{L} + n^{\da} \sigma_{R} \right) \label{eq:NL}\\
		 N(\varphi=\pi) &=& a \rho \ell \left(n^{\ua} \sigma_{\ua,R} + n^{\da} \sigma_{\da,R} \right)= a \rho \ell \left( n^{\ua} \sigma_{R} + n^{\da}\sigma_{L} \right)
		 \label{eq:NR}
\end{eqnarray}	
Here $n^\ua$ ($n^\da$) denotes the number of beam particles with
spin pointing upwards (downwards).
$\rho$ is the target density and $\ell$ its length.
$\sigma_L^\ua$ is the cross section for a beam particle with spin upwards scattered to the left. Equivalent definitions hold for the other cross sections appearing in equation~\ref{eq:NL} and~\ref{eq:NR}.
The second equal sign hold
because of the rotational symmetry:
\[
\sigma_{\da,L} = \sigma_{\ua,R} =: \sigma_R \quad \mbox{and} \quad 
 \sigma_{\da,R} =  \sigma_{\ua,L} =: \sigma_L
\]

 The beam polarization 
is defined as
\begin{equation}
 P = \frac{n^\ua - n^\da}{n^\ua + n^\da} \, ,
\end{equation}
and the analyzing power is given by
\begin{equation}
A = \frac{\sigma_{L} - \sigma_{R}}{\sigma_{L} + \sigma_{R}} \, .
\end{equation}

This allows one to write
 \begin{eqnarray*}
	N(\varphi=0) &=& a \rho \ell \left( n^\ua \sigma_R + n^\da \sigma_L \right) \\
	&=& a \rho l \frac{1}{2} \Big(  n^\ua \sigma_R + n^\da \sigma_L    +  n^\ua \sigma_R + n^\da \sigma_L  \Big)\\
	&=&  a\rho \ell \frac{1}{2} \Big(  n^\ua \sigma_R + { n^\ua \sigma_L}+{ n^\da \sigma_R}+ n^\da \sigma_L \\           \
	& &    \hspace{8mm}  +  n^\ua \sigma_R - {  n^\ua \sigma_L} - { n^\da \sigma_R}+ n^\da \sigma_L     \Big)     \\    
	&=&  a\rho \ell \frac{1}{2} \left(  (n^\ua+n^\da) (\sigma_R + \sigma_L)   +  (n^\ua-n^\da) (\sigma_R - \sigma_L)  \right) \\
	 &=&  a\rho \ell \underbrace{\frac{1}{2}  (\sigma_R + \sigma_L)}_{=\sigma}
	\left(  (n^\ua+n^\da)   +  (n^\ua-n^\da)\underbrace{\frac{\sigma_R - \sigma_L}{\sigma_R + \sigma_L}}_{=A} \right) \\
	&=&  a\rho \ell \underbrace{(n^\ua+n^\da)}_{=n} \sigma
	\left(  1   +  \underbrace{\frac{n^\ua-n^\da}{n^\ua+n^\da}}_{=P}  A \right)\\
	&=&  a \mathcal{L} \sigma \left(  1   +  P  A \right)  \, .   \\
\end{eqnarray*}
with the luminosity $\mathcal{L} = \rho \ell n$.
For the right side ($\varphi=\pi$) on finds correspondingly 
\begin{eqnarray*}
	N(\varphi=\pi)      &=&  a \mathcal{L} \sigma \left(  1   -  P  A \right)    \, . 
\end{eqnarray*}
Again using rotational symmetry and properly normalized we find for arbitrary azimuthal angles equations~\ref{eq:Npol} and ~\ref{eq:Nphi}:
\begin{equation}
\avrg{\frac{\mathrm{d} N(\varphi)}{\mathrm{d}\varphi}} = \frac{ a \mathcal{L} \sigma}{2 \pi} (1+A P \cos(\varphi)) \, .
\end{equation}

\section{Expectation values and averages}\label{app:avrg}
Note that there is a difference between the average denoted
be $\overline{(\dots)}$ and $\avrg{\dots}$ for odd powers 
of the cosine function.  
The expectation value is defined as
\begin{eqnarray}
\avrg{\cos(\varphi)^2} &=& \int_{acc} \cos(\varphi)^2 p(\varphi) \mathrm{d}\varphi  \nonumber\\
&=& \frac{1}{4\varphi_{max}} \, \int_{acc} \cos(\varphi)^2  \left(1+AP \cos(\varphi) \right) \mathrm{d}\varphi  \nonumber \\
&=& \frac{1}{4\varphi_{max}} \,\Big( \int_{-\varphi_{max }}^{\varphi_{max }} \cos(\varphi)^2  \left(1+AP \cos(\varphi) \right) \mathrm{d}\varphi  \nonumber\\
&& \, \,+\int_{\pi-\varphi_{max }}^{\pi+\varphi_{max }} \cos(\varphi)^2 \left(1+AP \cos(\varphi) \right) \mathrm{d}\varphi\, \Big) \nonumber\\
&=& \frac{\int_{-\varphi_{max }}^{\varphi_{max }} \cos(\varphi)^2\mathrm{d}\varphi}{2 \varphi_{max}} = \overline{\cos(\varphi)^2}
\end{eqnarray}
In general for even powers one obtains:
\begin{equation}
 \avrg{\cos(\varphi)^{2n}} = \overline{\cos(\varphi)^{2n}} \, , \quad n=0,1,2,\dots \, .
\end{equation}

For odd powers on the other hand we find
\begin{eqnarray}
\avrg{\cos(\varphi)} &=& \int_{acc} \cos(\varphi) p(\varphi) \mathrm{d}\varphi \nonumber\\
&=& \frac{1}{4\varphi_{max}} \, \int_{acc} \cos(\varphi)  \left(1+AP \cos(\varphi) \right) \mathrm{d}\varphi \nonumber \\
&=& \frac{1}{4\varphi_{max}} \,\Big( \int_{-\varphi_{max }}^{\varphi_{max }} \cos(\varphi)  \left(1+AP \cos(\varphi) \right) \mathrm{d}\varphi \nonumber \\
&& \quad \quad \, \,+\int_{\pi-\varphi_{max }}^{\pi+\varphi_{max }} \cos(\varphi) \left(1+AP \cos(\varphi) \right) \mathrm{d}\varphi\, \Big)\nonumber \\
&=& 0 + AP \avrg{\cos^2{\varphi}} \, .
\end{eqnarray}
In the evaluation of the FOM in equation~\ref{eq:fom_cnt} and~\ref{eq:fom_1}
\begin{equation}
\overline{\cos(\varphi)} = \frac{\int_{-\varphi_{max }}^{\varphi_{max }} \cos(\varphi)\mathrm{d}\varphi}{2 \varphi_{max}}
=\frac{-\int_{\pi-\varphi_{max}}^{\pi+\varphi_{max}} \cos(\varphi) \mathrm{d}\varphi}{2 \varphi_{max}}
\end{equation}
occurred. For odd powers one obtains
 \begin{equation}
 \avrg{\cos(\varphi)^{2n+1}} = AP  \avrg{\cos(\varphi)^{2n+2}} \ne \overline{\cos(\varphi)^{2n+1}} \, , \quad n=0,1,2,\dots \, .
 \end{equation}
Table~\ref{tab:expectation} list a few expectation values and averages used in the derivations.
\begin{table}
	\begin{center}
		\renewcommand*{\arraystretch}{1.5}
		\begin{tabular}{l|l|l}
			&   $w=\pm1$  &  $ w = \cos(\varphi)$ \\
			\hline 
			$\avrg{w}$ & $AP \, \overline{\cos(\varphi)} $ & $AP \, \avrg{\cos(\varphi)^2} $ \\
			$\avrg{w^2}$ & $   1  $            & $\avrg{\cos(\varphi)^2}$  \\
			$\avrg{wc}$ &    $\overline{\cos(\varphi)}$  &  $\avrg{\cos(\varphi)^2}$  \\
			$\avrg{w^2 c}$ &   $AP \, \avrg{\cos(\varphi)^2}$ & $AP\, \avrg{\cos(\varphi)^4}$   \\
			$\avrg{w^2 c^2}$ &  $\avrg{\cos(\varphi)^2}$ & $\avrg{\cos(\varphi)^4}$
		\end{tabular}
		\caption{Various expectation values for the two weighting factors. \label{tab:expectation}}
	\end{center}
\end{table}

\section{Error Propagation}\label{app:err}

Starting from the expression in equation~\ref{eq:Pw} ($w_i \equiv w(\varphi_i)$)
\begin{equation}\label{eq:Phat1}
\hat P = \frac{1}{A} \frac{\sum_i w_i}{\sum_i w_i \cos(\varphi_i)} \, ,
\end{equation}
we define 
\begin{eqnarray}
W &=& \sum_i w_i \quad \mbox{and}\\
V &=& \sum_i v_i = \sum_i w_i \cos(\varphi_i)  \, .
\end{eqnarray}
First we calculate the covariance matrix for $V$ and $W$:
\begin{eqnarray}
\mbox{Cov}(W,V) &=& \avrg{W V} - \avrg{W} \avrg{V} \\
  &=& \avrg{ \sum_i w_i \sum_j v_j} - \avrg{ \sum_i w_i} \avrg{\sum_j v_j}   \\
&=& \avrg{ \sum_{i=j} w_i v_i + \sum_{i\ne j} w_i v_j} 
- \avrg{ \sum_i w_i} \avrg{\sum_j v_j}  \\
&=& \avrg{N}  \avrg{W V} + \avrg{N(N-1)}  \avrg{W}  \avrg{V} - \avrg{N}^2 \avrg{ W} \avrg{V}  \\
&=& \avrg{N}  \avrg{WV} + \left(\avrg{N^2} - \avrg{N} - \avrg{N}^2\right)  \avrg{W}  \avrg{V}   \, .                    
\end{eqnarray}
Assuming a Poisson distribution for $N$, the term in parentheses vanishes.
Thus the elements of the covariance matrix of $W$ and $V$
are given by:
\begin{eqnarray*}
\mbox{cov}(W,V) &=& N \avrg{w^2 \cos(\varphi)} = \sum_i w_i^2 \cos(\varphi_i) \, , \\
\sigma_W^2 &=& N \avrg{w^2} = \sum w_i^2 \, , \\
\sigma_V^2 &=&N \avrg{w^2 \cos(\varphi)^2} = N \sum w_i^2 \cos(\varphi_i)^2 \, .
\end{eqnarray*}

Error propagation on equation~\ref{eq:Phat1} leads to
(using the shorthand notation $c = \cos(\varphi$))
\begin{eqnarray}
\sigma^2_{\hat P} &=& 
\left(
\frac{\partial \hat P}{\partial W} ,
\frac{\partial \hat P}{\partial V}
\right)
\left(
\begin{array}{cc}
\sigma_{W}^2  & \mbox{cov} \\
\mbox{cov}   & \sigma_{V}^2
\end{array}
\right)
\left(
\begin{array}{c}
\frac{\partial \hat P}{\partial W} \\
\frac{\partial \hat P}{\partial V}
\end{array}
\right)\nonumber \\
&=& \frac{N}{A^2}
\left(
\frac{1}{V} ,
-\frac{W}{V^2}
\right)
\left(
\begin{array}{cc}
\avrg{w^2}  & \avrg{w^2 c} \\
 \avrg{w^2 c}  & \avrg{w^2 c^2} \\
\end{array}
\right)
\left(
\begin{array}{c}
\frac{1}{V} \\
-\frac{W}{V^2}
\end{array}
\right) \nonumber \\
&=& \frac{1}{N A^2}\left( \frac{\avrg{w^2}}{\avrg{wc}^2}
- 2\frac{\avrg{w^2c}\avrg{w}}{\avrg{wc}^3}
 +\frac{\avrg{w^2 c^2}\avrg{w}^2}{\avrg{wc}^4}  \right)
\end{eqnarray}
In the last line we replaced $W=N\avrg{w}$ and $V=N\avrg{w c}$.

Given the properties of  $w$  ($w(\varphi)=-w(\pi+\varphi))$,
one finds according to table~\ref{tab:expectation}:
\begin{eqnarray}
  \avrg{w} &=& A P \avrg{w c} \, ,\\
  \avrg{w^2 c} &=& A P \avrg{w^2 c^2} \, .
\end{eqnarray}
This leads to
\begin{equation}
 \mbox{FOM} = N A^2 \, \frac{\avrg{wc}^2}{\avrg{w^2}} \left( 1 - (AP)^2 \frac{\avrg{w^2 c^2}}{\avrg{w^2}} \right)^{-1} \, .
\end{equation}

Finally for the FOMs read
\begin{eqnarray}
 \mbox{FOM}^{\pm 1} &=& N A^2 \, \overline{\cos(\varphi)}^2\left( 1 - (AP)^2 \, \overline{\cos(\varphi)^2}\right)^{-1}  \, , \label{eq:fom_1_app}\\
  \mbox{FOM}^{\cos(\varphi)} &=& N A^2 
  \, \overline{\cos(\varphi)^2}\left( 1 - (AP)^2 \, \frac{\overline{\cos(\varphi)^4}}{\overline{\cos(\varphi)^2}}\right)^{-1}  \, .
  \label{eq:fom_c_app}
\end{eqnarray}

To leading order equations~\ref{eq:fom_1_app} and \ref{eq:fom_c_app} agree with the results given in equations~\ref{eq:fom_1} and \ref{eq:fom_c}.
Moreover the FOM for the case $w=\cos(\varphi)$ coincides
with the FOM derived for the maximum likelihood method in equation~\ref{eq:fom_ml}.

As explained in section~\ref{sec:weight}
the estimator 
\[
 \hat P = \frac{1}{\overline{\cos(\varphi)}} \frac{N_L - N_R}{N_L + N_R}
\]
leads to 
\[
\mbox{FOM} = N A^2 \, \overline{\cos(\varphi)}^2 \, \left( 1 - (AP) ^2 \, \overline{\cos(\varphi)^2}\right)^{-1}  \, ,
\]
if $\overline{\cos(\varphi)}$ is evaluated from a sum over events. 

In case $\overline{\cos(\varphi)}$ is supposed to be known, e.g.
from the integral $\int{\cos(\varphi)} \mathrm{d}\varphi$, the FOM is slightly different
at order $(AP)^2$ where now a factor $\overline{\cos(\varphi)}^2$
instead of $\overline{\cos(\varphi)^2}$ appears.
\[
  \mbox{FOM} = N A^2 \, \overline{\cos(\varphi)}^2 \, \left( 1 - (AP) ^2 \overline{\cos(\varphi)}^2\right)^{-1} \, .
\]

\bibliography{/home/pretz/bibtex/literature_axion.bib,/home/pretz/bibtex/literature_edm.bib,/home/pretz/bibtex/statistics.bib,/home/pretz/bibtex/literature_dis.bib}

\bibliographystyle{ieeetr}



\end{document}